\documentclass[aps,prl,twocolumn,nofootinbib,twoside,floatfix]{revtex4}

\usepackage{amsmath}
\usepackage{amssymb}
\usepackage{graphicx}
\usepackage{caption}

\newcommand{\Cal}[1]{\ensuremath{\mathcal{#1}}}

\newcommand{\fig}[1]{Fig. \ref{#1}}
\newcommand{\figs}[1]{Figs. \ref{#1}}

\newcommand{\ph}[1]{\phantom{#1}}

\newcommand{\p}{\ensuremath{\partial}}

\newcommand{\avg}[1]{\ensuremath{\langle #1 \rangle}}

\newcommand{\be}{\begin{equation}}
\newcommand{\ee}{\end{equation}}

\newcommand{\ti}[1]{{\tilde #1}}

\newcommand{\vphi}{\varphi}
\newcommand{\Mbar}{\ensuremath{\bar{\Cal{M}}}}

\begin{document}
\title{Cosmic Inhomogeneities and the Average Cosmological Dynamics}
\author{Aseem Paranjape}
\email{aseem@tifr.res.in}

\author{T. P. Singh}
\affiliation{Tata Institute of Fundamental Research, Homi Bhabha
Road, Mumbai - 400 005, India\\}

\begin{abstract}
\noindent
If 
general relativity (GR) describes 
the 
expansion of the 
Universe, 
the observed
cosmic acceleration implies 
the existence of a 
`dark energy'.
However, 
while the Universe is 
on average 
homogeneous on
large scales, it is inhomogeneous on smaller
scales. 
While GR governs 
the 
dynamics of the \emph{in}homogeneous
Universe, 
the
averaged 
\emph{homogeneous} Universe 
obeys modified Einstein equations. 
Can such 
modifications 
alone 
explain the acceleration? 
For a simple generic model with realistic initial conditions, we  
show the answer to be `no'. Averaging effects negligibly influence 
the cosmological dynamics. 
%
\end{abstract}

\maketitle
Cosmological observations 
have established that
we live in an accelerating Universe \cite{accn}. 
Assuming 
the dynamics of the expanding Universe 
to be 
described by 
general relativity
(GR)
, the observed
acceleration compels us to conclude 
the existence of a `substance' with negative pressure named 
`dark energy' \cite{sami}. The standard model of
particle physics offers no candidate for dark energy, nor does it
provide any convincing clues as to its origin. While this situation
might suggest new physics, it is still worthwhile to ask if a
conventional explanation for the acceleration might yet be possible. 

It is 
true that the standard hot Big Bang model 
for 
a \textit{homogeneous and isotropic} Universe 
described by the Friedmann-Lema\^itre-Roberton-Walker (FLRW) solution 
of Einstein's equations agrees well with observations. However, the 
actual Universe is highly {\it inhomogeneous} and homogeneity is
recovered only in an average sense on sufficiently large
scales \cite{homog}. Strictly speaking, 
only the inhomogeneous Universe 
obeys 
Einstein's 
equations,
and recovering the homogeneous limit by averaging leads to corrections
in the equations 
\cite{ellis}. Could 
these correction terms (backreaction) 
be significant enough to account for dark energy? 
This question has attracted considerable %
 recent 
attention 
\cite{corre,buchert,kolb}. 

One might 
simply 
take 
the view that 
since the 
perturbed FLRW model
works well, 
backreaction 
from averaging these small perturbations cannot be large. There is
however a subtle catch here. 
The evolution of 
inhomogeneities depends on 
the background FLRW solution, which in turn 
depends on the backreaction. 
One cannot \emph{a priori} rule
out a runaway situation in which the 
backreaction is 
fed by and 
reinforces 
a strong 
evolution of the
inhomogeneities, perhaps leading to a dark energy like component at
late times. 

A covariant averaging formalism for 
GR 
known as
macroscopic gravity (MG)
has been developed by Zalaletdinov
\cite{zala}. In MG
the connection on the inhomogeneous manifold 
\Cal{M}\ 
is
averaged to obtain a connection on a new ``averaged manifold'' 
\Mbar\ 
in a
rigorous manner. As a consequence, the averaging of 
Einstein's  
equations for 
\Cal{M}\ 
leads to
a set of corrected equations for 
\Mbar\ 
with the corrections being quadratic in the underlying
connection. 
Of relevance 
for 
cosmology is the special case where averaging is performed on specific
spatial slices; the 
backreaction 
for this case 
was 
obtained in \cite{spatavglim}.  
Further, 
\emph{if} it can be self-consistently established that
the perturbed FLRW model works well throughout the evolution of the
Universe, then 
the 
backreaction 
must 
be evaluated for this specific
perturbative case. This has been done in \cite{pertbakrxn}, by
assuming the metric to have the form   
\be
ds^2 = a^2(\eta)\left[ -(1+2\vphi)d\eta^2 +
  (1-2\psi)\gamma_{AB}dx^Adx^B \right]\,, 
\label{2eq1}
\ee
with $\eta$ the conformal time and $\gamma_{AB}$ the (flat) $3$-space
metric. In the present paper we work with the
cosmic time $\tau$ related to the conformal time by $d\tau =
a(\eta)d\eta$, 
with appropriate conversions applied. 
The modified 
FLRW 
equations are given by  
\begin{align}
&H^2 \equiv \left(\frac{1}{a}\frac{da}{d\tau}\right)^2 = \frac{8\pi
    G_N}{3}\bar\rho -  \frac{1}{6}\left[ \Cal{P}^{(1)} + \Cal{S}^{(1)}
    \right]  \,, 
\nonumber\\ 
&\frac{1}{a}\frac{d^2a}{d\tau^2} = -\frac{4\pi
  G_N}{3}\left(\bar\rho+3\bar p\right) + \frac{1}{3}\left[
  \Cal{P}^{(1)} + \Cal{P}^{(2)} +
  \Cal{S}^{(2)} \right] \,,
\label{2eq2}
\end{align}
where $\bar\rho$ and $\bar p$ are the %
background 
energy density and pressure respectively 
and the
combinations $(\Cal{P}^{(1)} + \Cal{S}^{(1)})$ and $(\Cal{P}^{(1)} +
\Cal{P}^{(2)} +  \Cal{S}^{(2)})$ are 
covariant scalars given
by 
\begin{align}
&\Cal{P}^{(1)} = \bigg[ \, 2\avg{(\p_\tau\psi)^2} + 
    \avg{\left(\p_\tau\vphi - \p_\tau\psi\right)^2} \nonumber\\
&\ph{\Cal{P}^{(1)} = \frac{1}{a^2}}    -
    \avg{\left(\nabla_A\nabla_B\p_\tau\beta\right)
      \left(\nabla^A\nabla^B\p_\tau\beta\right)} \,  \bigg]
  \,, \nonumber\\   
&\Cal{S}^{(1)} = -\frac{1}{a^2} \bigg[ 6\avg{\p_A\psi\p^A\psi} +
    \avg{\p_A(\vphi-\psi)\p^A(\vphi-\psi)}  \nonumber\\
&\ph{\Cal{S}^{(1)}= -\frac{1}{a^2}}  -
    \avg{(\nabla_A\nabla_B\nabla_C\beta)
      (\nabla^A\nabla^B\nabla^C\beta)}  \bigg]  
  \,,\nonumber\\ 
&\Cal{P}^{(1)} + \Cal{P}^{(2)} = \bigg[
    \avg{\p_\tau\vphi(\p_\tau\vphi-\p_\tau\psi)}  \nonumber\\    
&\ph{\Cal{P}^{(1)}}- 2H\left\{\,
    \avg{\vphi\p_\tau\vphi} -  \avg{\psi\p_\tau\psi}  +
    \avg{(\vphi-\psi)\p_\tau\psi}  \right.  \nonumber\\   
&\ph{\Cal{P}^{(1)}} \left.  +  
    \avg{\psi\p_\tau(\vphi-\psi)}  +
    \avg{(\nabla_A\nabla_B\beta)(\nabla^A\nabla^B\p_\tau\beta)}\, 
    \right\}  \bigg]\,, \nonumber\\  
&\Cal{S}^{(2)} = -\bigg[
    \avg{\p^A(\p_\tau^2\beta + H\p_\tau\beta) \p_A\left(\vphi -
      a^2H\p_\tau\beta \right)}  \bigg] 
  \,. \label{corrscal}
\end{align}
Here $\nabla_A$ ($A,B=1,2,3$) is the $3$-space covariant derivative 
and 
$\beta$ 
solves 
$\nabla^2\beta = \vphi - 3\psi$ with 
$\nabla^2\equiv\gamma^{AB}\nabla_A\nabla_B$, 
%
with 
the condition that %
when $\vphi=0=\psi$, $\beta=0$. 
The angular brackets in \eqref{corrscal} denote a spatial
averaging defined
as
\be
\avg{f}(\tau,\vec{x}) = \frac{1}{V_L}\int_{\Cal{V}(\vec{x})}{d^3y 
   f(\tau,\vec{y})} \,, 
\label{2eq5}
\ee
for any function $f(\tau,\vec{x})$ where $\Cal{V}(\vec{x})$ is the
$3$-dimensional averaging domain 
of ``Eulerian'' 
length scale $L$ (comoving with the background) and 
volume $V_L$. This averaging operation is derived by considering a
spatial averaging limit of the full 
MG 
averaging technique
and working in a specific ``volume preserving'' gauge (in which the
metric determinant 
depends 
only 
on 
cosmic time) with the form  
\be
ds^2 = -(1+6\ti{\psi})d\tau^2 + a^2(\tau)(1-2\ti{\psi})d\vec{x}^2\,,
\label{vpgauge}
\ee
where $\ti{\psi}=\vphi/3$
(see 
\cite{pertbakrxn}). 
This procedure is 
thus 
different from the spatial averaging of Buchert
\cite{buchert,spatavglim}. Having defined the averaging in this gauge,
one 
rewrites 
all quantities in terms of the conformal Newtonian
gauge potentials $\vphi$ and $\psi$, and \eqref{corrscal} show the
final result.  

In 
this paper we work
with the background metric in 
spherical coordinates,
\be
ds^2_{bg} = -d\tau^2 + a^2(\tau)\left[d\ti{r}^2 + \ti{r}^2d\Omega^2
  \right] \,,  
\label{2eq6}
\ee
where $\ti{r}$ is the 
Eulerian 
radial coordinate. 
It is then most
convenient to consider a single domain which is a sphere of radius $L$
centered at the origin. The expressions for the 
backreaction \eqref{corrscal} will then correspond to this single
domain. 

When nonlinear structures 
such as
galaxy 
clusters
and voids form during 
late stages of 
evolution of the Universe, 
is 
the spacetime 
metric 
still 
perturbed FLRW? 
The answer 
is yes, provided matter peculiar velocities remain
small. We showed this by considering a simple but generic model of
spherically symmetric pressureless dust collapse \cite{sphcoll} (see
also \cite{karel}). We 
now apply the backreaction results
\eqref{corrscal} to our 
collapse model and show that the
corrections are extremely small. This is the first example of a fully
covariant calculation of the backreaction valid even in the
nonlinear regime of structure formation. 

The spacetime of the spherically collapsing dust 
is described by
the Lema\^itre-Tolman-Bondi (LTB) metric given by
\be
ds^2 = -dt^2 + \frac{R^{\prime2}dr^2}{1-k(r)r^2} + R^2d\Omega^2 \,.
\label{3eq1}
\ee
Here $t$ is the proper time measured by observers with fixed
coordinate $r$, which 
comoves 
with the dust. $R(t,r)$ is the
area radius of the dust shell labelled by $r$, and satisfies
the equation 
${\dot R}^2 = 2GM/R - kr^2$. 
Here $M(r)$ is the mass 
inside each comoving shell
and a dot denotes a derivative with respect to 
$t$. The
energy density of dust measured by 
a comoving observer 
satisfies 
$\rho(t,r) = M^\prime/4\pi R^2R^\prime$ 
where the prime denotes a derivative with respect to 
$r$. Initial conditions are completely specified by choosing 
an initial density $\rho(t_i,r)$, velocity $\dot R(t_i,r)$ 
and 
an 
initial scaling 
function 
$R(t_i,r)$. In \cite{sphcoll} we made the
following choices: 
To set the initial conditions as being a
perturbation around the Einstein-deSitter (EdS) solution characterised
by the scale factor
$a(t) \equiv (t/t_0)^{2/3}$
with $t_0=2/(3H_0)$ where $H_0$ is the standard Hubble constant, we
chose the functions
\begin{align}
&R(t_i,r)=a_ir ~~;~~ \dot R(t_i,r) = a_iH_ir\,,
\label{3eq5}\\
&\rho(t_i,r)=\bar\rho_{i}\left\{
\begin{array}{l}
(1+\delta_\ast),\,~~~r<r_\ast\\
(1-\delta_v),\,~~~r_\ast<r<r_v\\
\,1,~~~~~~~~~~~~r>r_v\,,
\end{array}\right .
\label{3eq6}
\end{align}
where $\bar\rho_{i}=\bar\rho(t_i)$, $a_i = (t_i/t_0)^{2/3}$,
$H_i=2/(3t_i)$, and $\delta_\ast$, $\delta_v$, $r_\ast$ and $r_v$ are
constants whose values were chosen so that the system being described
is initially a small overdensity of extent $0<r<r_\ast$, surrounded by
a small underdensity out to radius $r_v$. In particular, the following
values were chosen for the various parameters (cf. Table 1 of
\cite{sphcoll}) 
\begin{align}
\nonumber
&a_i=0.001~~;~~ H_0 = 72\, {\rm km\, s}^{-1}{\rm Mpc}^{-1}\,, \nonumber\\
&\delta_\ast=2.21\times10^{-3}~~;~~\delta_v=5\times10^{-3}\,,
  \nonumber\\ 
&r_\ast=16.7\, {\rm Mpc}~~;~~ r_v=23.5\, {\rm Mpc}\,.
\end{align}
The exact solution for $R(t,r)$ can be written in parametric form in
terms of trigonometric or hyperbolic functions depending on the sign
of $k(r)$. One then numerically obtains
the function $R(t,r)$ in the 
region of interest in the $(t,r)$ plane.

It was further shown in \cite{sphcoll} that for the chosen model, for
all times $t_i<t<t_0$, the LTB metric \eqref{3eq1} can be brought to
the form 
\be
ds^2 = -(1+2\vphi)d\tau^2 + a^2(\tau)(1-2\psi)(d\ti{r}^2 +
\ti{r}^2d\Omega^2) \,,
\label{3eq8}
\ee
where $\vphi$ and $\psi$ satisfy $|\vphi|,|\psi|\ll1$, and
$a(\tau)=(3H_0\tau/2)^{2/3}$. This is achieved by the coordinate
transformation $(t,r)\to(\tau,\ti{r})$ given by 
\be
\nonumber
\tau = t + \xi^0(t,r) ~~;~~ \ti{r} = \left(R(t,r)/a(t)\right) \left( 1
+ \xi(t,r) \right)\,.
\ee
Here $\xi^0$ and $\xi$ are assumed to satisfy $|\xi^0H|,|\xi|\ll1$ and
are determined by integrating the equations
\be
\nonumber
\xi^\prime = (1/2)\left(k(r)r^2+(a\ti{v})^2\right) \left(
R^\prime/R \right) ~~;~~ 
\xi^{0\prime} = a\ti{v}R^\prime
\ee
where $\ti{v}\equiv(\p\ti{r}/\p t)\approx \p_t(R/a)$ is the comoving
peculiar velocity, also assumed to remain small \cite{footnote}. This
is a self-consistent calculation, in 
that one assumes 
such functions 
to 
exist 
to set up the equations, and then
solves for them showing that they satisfy the required
properties. In particular, one finds that the peculiar velocity
$\ti{v}$ does remain small throughout the evolution. 
Of course, models with large peculiar velocities can be studied
(e.g. the $\sim1$Gpc void studied in \cite{kolb}) 
and such a case 
would indicate a breakdown of the weak field
approximation. What is important however, is whether such
inhomogeneities are generic. The parameter choices 
in our model 
reflect the nature of typical
inhomogeneities at the Last Scattering epoch
and do not lead to
large underdense regions and/or large peculiar
velocities. 
With these choices, we find that the metric 
potentials $\vphi$ and $\psi$ can be obtained self-consistently as   
$\vphi = -\dot\xi^0 + \left(a\ti{v} \right)^2/2$ and 
$\psi = \xi^0H + \xi$, 
where, at the leading order
we have $H \equiv 2/3t \approx 2/3\tau$. It can be analytically shown
by working in the Newtonian gauge \cite{karel} that the metric
potentials $\vphi$, $\psi$ are in fact equal at leading order, and
this can also be checked explicitly. We emphasize that the metric
potentials remain small in 
magnitude compared to unity, even though the density contrast
$(\rho/\bar\rho-1)$ becomes completely nonlinear at late times
\cite{sphcoll}. While a more careful 
second order 
calculation can
in principle
be
performed, 
it is not difficult to show that 
the 
higher order terms will
contribute insignificantly to the backreaction, so long as peculiar
velocities are small. [This stems from the basic structure of the
  nonlinear 
  metric potentials which depends on an expansion in $(HR)\ll1$.] 
Hereon 
when considering time derivatives of small quantities we
shall not distinguish between $t$ and $\tau$. Terms quadratic in
$\ti{v}$ 
are retained since on dimensional grounds one
expects $a\ti{v}\sim HR$ whereas $\vphi,\psi\sim(HR)^2$, which is
confirmed by 
our numerics 
(see also \cite{karel}).

We 
now compute the backreaction 
given in
\eqref{corrscal}. These expressions 
were derived
in \cite{pertbakrxn} under the requirement that the averaging
operation be free of gauge related ambiguities, in \emph{linear}
perturbation theory. However, the actual conditions used to derive
\eqref{corrscal} only depended 
on considering 
leading order effects in the \emph{metric} perturbations. 
A key step was the transformation between the metric
\eqref{3eq8} (in Cartesian spatial coordinates)
and the volume
preserving form \eqref{vpgauge}, which was achieved by the
transformation $\tau\to\tau$, $x^A\to x^A+\p^A\beta$ ($A=1,2,3$), 
where $\beta$ is the function appearing in \eqref{corrscal}. In the
present context, the same transformation remains valid \emph{at the
  leading order}
and hence the 
backreaction 
in \eqref{corrscal} 
is 
physically relevant here as well. We emphasize that
this truncated averaging operation remains valid even at late times
since the weak field approximation for gravity works well during 
nonlinear 
structure formation.

Since our numerical results 
involve 
$(t,r)$
where $r$ 
comoves 
with the matter, 
we 
must 
reexpress the averaging operation \eqref{2eq5}
in terms of these variables. It is easy to show that, at the leading
order, the average of a scalar $s(t,r)$ defined in \eqref{2eq5} can be
written as 
\be
\avg{s} = \frac{3}{(a(t)L)^3} \int_0^{r_L(t)}{sR^2R^\prime dr} \,,
\label{4eq2}
\ee
where 
$r_L(t)$ solves 
$R(t,r_L(t)) = a(t)L$.
Recall that 
\eqref{4eq2} gives the average of $s$ over a
single domain centered at the origin, which is what we 
restrict ourselves to in this paper. 
We are constrained to consider values $r<r_v$
due to unphysical shell crossing singularities in the region
beyond
(see \cite{sphcoll}), 
so 
the 
largest value of $L$ 
we can choose is $L=r_v$, which then
ensures $r_L(t)<r_v$ since $r_L(t)$ is a decreasing function for this
choice. This gives us an 
Eulerian 
averaging scale of $L=23.5$ Mpc 
which is smaller than the more realistic
expected value of $\sim100h^{-1}$Mpc. 
Our model potentials $\vphi$, $\psi$ and their derivatives hence do
not strictly average to zero as needed 
\cite{sphcoll,pertbakrxn}. One can check 
that 
actual average values 
$\avg{(\p_A\vphi)}^2$ are small ($\lesssim10\%$) for all times
compared to terms like $\avg{(\p_A\vphi)^2}$ which are needed in the
backreaction calculations, 
although it turns out that the \emph{time} derivatives satisfy
$\avg{\dot\vphi}^2\sim\avg{(\dot\vphi)^2}$. 
However, since the averaging scale chosen here is large
enough to encompass all the inhomogeneity of \emph{this} system, we
expect that our estimates for the backreaction 
are fairly representative.

Consider now the function $\beta$
which 
satisfies the Poisson equation on a flat $3$-space background, 
such that $\beta=0$ if $\vphi=0=\psi$. 
We can directly write the solution for $\beta$ in terms of
the 
coordinate $\ti{r}$ as,
\begin{align}
\nonumber
\beta(\tau,\ti{r}) &= -(1/4\pi) \int{ d^3y
  q(\tau,\vec{y})/|\vec{\ti{r}}-\vec{y}| }
\nonumber\\ 
& = -\frac{1}{\ti{r}}\int_0^{\ti{r}}{q\, y^2dy} -
\int_{\ti{r}}^\infty{q\, ydy}\,,
\end{align}
where $q \equiv -2\vphi$ (we have set $\vphi=\psi$ at leading order)
and the integration is over 
Eulerian 
spatial coordinates. 
The following relations 
are 
useful in the calculations,  
\begin{subequations}
\begin{align}
\p_{\ti{r}}\beta &= \frac{1}{\ti{r}^2} \int_0^{\ti{r}}{
  q\,y^2dy}\,, \label{5eq3a}\\ 
\p_{\ti{r}}^2\beta &= q -
\frac{2}{\ti{r}}\p_{\ti{r}}\beta\,, \label{5eq3b} \\
\p_{\ti{r}}^3\beta &= \frac{6}{\ti{r}^2}\p_{\ti{r}}\beta -
\frac{2}{\ti{r}}q + \p_{\ti{r}}q\,. \label{5eq3c}
\end{align}
\label{5eq3}
\end{subequations}
We will need 
$\p_{\ti{r}}\beta$, $\p_{\ti{r}}^2\beta$
and $\p_{\ti{r}}^3\beta$ as functions of 
$(t,r)$,
which 
is 
done by replacing $\tau$ and $\ti{r}$ at leading order by
$t$ and 
$R/a$ 
respectively. This gives us 
\begin{subequations}
\begin{align}
(\p_{\ti{r}}\beta)(t,r) &= \frac{1}{aR^2} \int_0^{r}{
  q\,R^2R^\prime dr}\,, \label{5eq4a}\\ 
(\p_{\ti{r}}^2\beta)(t,r) &= q -
\frac{2}{R^3}\int_0^{r}{ q\,R^2R^\prime dr}\,, \label{5eq4b} \\ 
(\p_{\ti{r}}^3\beta)(t,r) &= \frac{6a}{R^4}\int_0^{r}{q\,R^2R^\prime
  dr} - \frac{2a}{R}q + \frac{a}{R^\prime}q^\prime\,, \label{5eq4c}
\end{align}
\label{5eq4}
\end{subequations}
where in the last equation we have used 
$\p_{\ti{r}}q = (a/R^\prime)q^\prime$
at leading order. 

\begin{figure}[t]
\includegraphics[width=8.6cm]{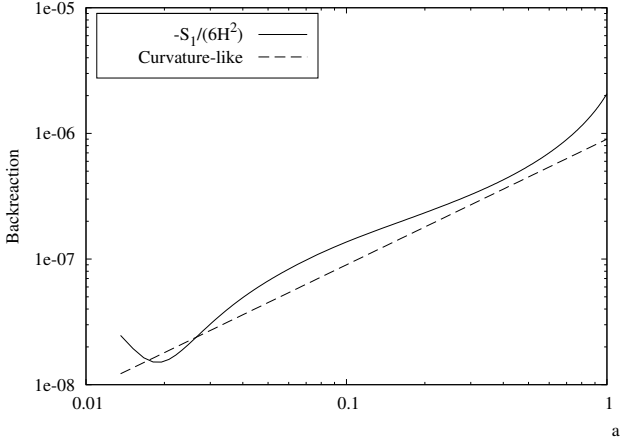}
\caption{\small The evolution of $-\Cal{S}^{(1)}/6H^2$.
  Also shown is a hypothetical curvature-like correction, evolving
  like $\sim a^{-2}$.}   
\label{fig1}
\end{figure}

Also, noting that the time derivatives in \eqref{corrscal} are taken
keeping 
$\ti{r}$ fixed, we have 
\begin{subequations}
\begin{align}
(\p_{\ti{r}}\dot\beta)(t,r) &= \frac{1}{aR^2} \int_0^{r}{
  \dot q\,R^2R^\prime dr}\,, \label{5eq5a}\\ 
(\p_{\ti{r}}{\ddot\beta})(t,r) &= \frac{1}{aR^2} \int_0^{r}{
  \ddot q\,R^2R^\prime dr}\,, \label{5eq5b} \\ 
(\p_{\ti{r}}^2\dot\beta)(t,r) &= \dot q - \frac{2}{R^3}\int_0^{r}{ 
  \dot q\,R^2R^\prime dr}
  \,, \label{5eq5c} 
\end{align}
\label{5eq5}
\end{subequations}
which follow from \eqref{5eq3}. The expressions 
\eqref{corrscal}, rewritten in terms of 
$t$ and valid at leading order, 
reduce to
\begin{align}
&\Cal{P}^{(1)} =  \bigg[ \, 2\avg{(\dot\vphi)^2} 
-    \avg{(\p_{\ti{r}}^2\dot\beta)^2} -
    2\avg{(1/\ti{r}^2)(\p_{\ti{r}}\dot\beta)^2} \,  \bigg] 
  \,, \nonumber\\
&\Cal{S}^{(1)} = -\frac{1}{a^2} \bigg[ 6\avg{(\p_{\ti{r}}\vphi)^2} 
  -  \avg{(\p_{\ti{r}}^3\beta)^2}
- 6\avg{(\p_{\ti{r}}\beta
      - \ti{r}\p_{\ti{r}}^2\beta)^2/\ti{r}^4}  \bigg]  
  \,,\nonumber\\ 
&\Cal{P}^{(1)} + \Cal{P}^{(2)} = - 2H\bigg[ 
    \avg{(\p_{\ti{r}}^2\beta)(\p_{\ti{r}}^2\dot\beta)} +
    2\avg{(1/\ti{r}^2)(\p_{\ti{r}}\beta)(\p_{\ti{r}}\dot\beta)} \,  
 \bigg]\,, \nonumber\\  
&\Cal{S}^{(2)} = \avg{(\p_{\ti{r}}{\ddot\beta} +
    H\p_{\ti{r}}{\dot\beta})(a^2H\p_{\ti{r}}{\dot\beta} -
    \p_{\ti{r}}\vphi)}  
  \,, \label{5eq6}
\end{align}
where the angular brackets are now defined by \eqref{4eq2} and the
various integrands can be read off using \eqref{5eq4}, \eqref{5eq5} and
the results $\p_{\ti{r}}\vphi \approx (a/R^\prime)\vphi^\prime$ and
$\ti{r}\approx (R/a)$. 

\begin{figure}[t]
\includegraphics[width=8.6cm]{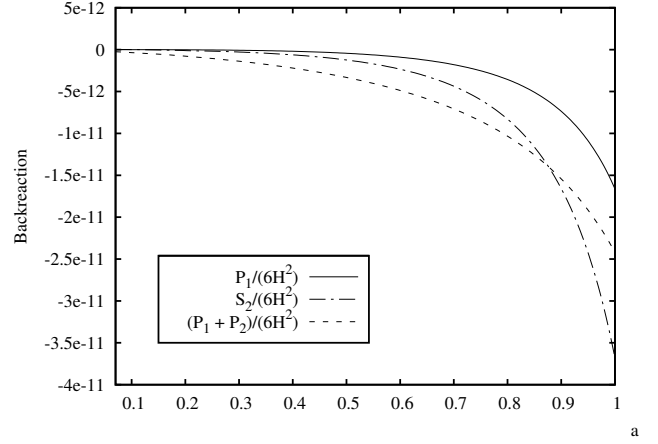}
\caption{\small The normalised evolution of the backreaction functions
  other than $\Cal{S}^{(1)}$. 
To enhance contrast, a strongly decaying
  early time mode for $\Cal{P}^{(1)}/H^2$ 
has not been shown. }
\label{fig2}
\end{figure}

\figs{fig1} and \ref{fig2} show results of numerical calculations
performed with {\it Mathematica}. 
\fig{fig1} shows the evolution of the dominant correction
$-\Cal{S}^{(1)}/6H^2$, 
as a function
of the scale factor. The 
dashed 
line shows a hypothetical curvature like correction. 
The actual backreaction evolves differently 
due to significant evolution of $\vphi$. Note that the largest value
of $|\Cal{S}^{(1)}/H^2|$ computed here is $\sim10^{-6}$, whereas
estimates using \emph{linear} theory suggest 
this value should be 
$\sim10^{-4}$ \cite{pertbakrxn}. This
discrepancy highlights an issue noted
in \cite{pertbakrxn}, namely that nonlinear inhomogeneities on small
scales \emph{do not contribute significantly} to the backreaction. Our
model has no large scale inhomogeneities and underestimates the
backreaction. Reassuringly, accounting for the deficit only requires a
calculation in \emph{linear} theory, such as the one in
\cite{pertbakrxn}. \fig{fig2} shows the evolution of the remaining %
(normalised) 
integrals. 
An initial rapid decay of $\Cal{P}^{(1)}/H^2$ starting from 
$\sim10^{-8}$ has not been shown, in order to enhance the
contrast in the late time behaviour of the three functions. The other 
functions remain subdominant compared to $\Cal{P}^{(1)}$ at the early
times not shown.

Our covariant and self-consistent calculation of the backreaction in
this spherical collapse model establishes that inhomogeneities have an
insignificant impact on the average cosmological dynamics. In
particular, the observed cosmic acceleration cannot be explained by
the averaging of inhomogeneities. Our nonlinear dust model can be
regarded as representing a realistic situation, because it has a
overdensity-void structure, and departure from sphericity, tidal
interactions, and second order corrections are not expected to
introduce any significant change in the results. What appears true in
general is that as long as peculiar velocities remain small, as seems
to be the case in the real Universe, a description as a perturbed FLRW
model is valid, and this keeps the back-reaction small.


\begin{thebibliography}{10}
\bibitem{footnote}
This corrects an error in Eq.35 of \cite{sphcoll}. We 
thank 
Karel Van Acoleyen for pointing this out to us.
\end{thebibliography}
\end{document}